\documentclass[10pt,a4paper,twocolumn,aps,prl,notitlepage,nofootinbib,floatfix]{revtex4-1}
\usepackage[T1]{fontenc}
\usepackage[latin2]{inputenc}
\usepackage{amsmath}
\usepackage{esint}

\makeatletter

\pdfpageheight\paperheight
\pdfpagewidth\paperwidth

\usepackage{graphicx}
\usepackage{color}
\usepackage[normalem]{ulem}

\makeatother

\begin{document}

\title{Gravitational waves in Friedman-Lemaitre-Robertson-Walker cosmology,
material perturbations and cosmological rotation,  and the Huygens principle}

\author{Wojciech Kulczycki}

\author{Edward Malec}

\affiliation{Instytut Fizyki im.\ Mariana Smoluchowskiego, Uniwersytet Jagiello\'nski,
\L ojasiewicza 11, 30-348 Krak\'ow, Poland}
\begin{abstract}
We analyze propagation equations for the polar modes of gravitational
waves in cosmological space-times. We prove that polar gravitational
waves must perturb the density and non-azimuthal  components of the velocity  of material medium
of the Friedman-Lemaitre-Robertson-Walker spacetimes. Axial gravitational waves can influence
only the azimuthal velocity, leading to local cosmological rotation. The whole gravitational
dynamics reduces to the single \textquotedbl{}master equation\textquotedbl{}
that has the same form    for polar and axial modes. That allows us to conclude
that the status of the Huygens principle is the same for axial and polar
gravitational waves. In particular, this principle is valid exactly
in radiation spacetimes with the vanishing cosmological constant,
and it is broken otherwise. 
\end{abstract}
\maketitle

\section{Introduction}

The canonical way to establish the topological type of the Universe
is through the analysis of the Hubble relation \cite{Weinberg2008}.
There would seem to exist an alternative --- it is known that gravitational
waves can backscatter on the curvature of the spacetime. The effects
of the backscatter (or equivalently, the breakdown of the Huygens
principle) would give, in principle at least, a new local way to see
the global topology of a cosmological spacetime in a way analogous
to effects known in the Schwarzschild spacetime. Quasinormal modes
and tails ( \cite{Vishveshwara}, \cite{Price}) that are present
in the radiation that originated beneath the light sphere of a black
hole would allow its identification. These effects can be vigorous,
as seen in the recent detection of gravitational waves \cite{Abbott}.
They have been discussed in theoretical studies of \cite{KMS} and
\cite{KMRS}; the latter have shown a positive correlation between
the strength of backscattering and the intensity of ringing. The results
of \cite{MW} have shown that while axial modes of gravitational waves
indeed \textbf{do backscatter} (with the exception of purely radiative
spacetimes), but this effect becomes significant only for length-waves
comparable to the Hubble radius, and thus cannot be seen in local
observations.

The original motivation of  this paper was to complete the analysis of \cite{MW} by discussing
polar modes. That constitutes a technically demanding task, but the
conclusion is the same as in \cite{MW}: backscattering effects cannot
be seen locally.  There is, however, a feature, that (we think) is
more interesting: it appears that polar gravitational waves must perturb
the material content --- both its density and poloidal components of the velocity ---  of the cosmological universe. 
This should be contrasted with the case of axial gravitational waves, 
where  one  may choose initial data in such a way, that gravitational waves completely  
decouple from matter. The new result is   that   the azimuthal  component of the
velocity --- and only azimuthal one --- of the background fluid can be affected; moreover, it is expected to be
 perturbed by generic initial data.
 Thus axial modes can lead to a local (but  cosmologically significant)  rotation.

Regge and Wheeler extracted two gauge-independent linearized modes
(axial and polar) of gravitational waves in the Schwarzschild spacetime
(\cite{R-W}) and derived an evolution equation for axial modes. They
made minor mistakes in equations describing propagation of polar modes,
corrected by Zerilli \cite{Zerilli}. In the cosmological context,
Malec and Wyl\c{e}\.zek \cite{MW} adapted the method of Regge and Wheeler
and obtained a propagation equation for axial modes in Friedman-Lemaitre-Robertson-Walker
(FLRW thereafter) spacetimes, with vanishing cosmological constant.
They found that the Huygens principle is strictly obeyed in radiation
spacetimes and approximately --- for relatively short length-waves
- in spacetimes filled by dust.

The content of this paper is following. Next section specifies the
FRLW geometries with a non-zero cosmological constant, and their (Regge-Wheeler)
perturbations. The material is composed of perfect fluids, and we
write down both the relevant energy-momentum tensor, and its linearization.

Section 3 supplements formerly known results obtained in \cite{MW}. It
appears that matter density cannot be perturbed by travelling axial gravitational
waves, and the description of waves reduces to a simple propagation
equation ( a \textquotedbl{}master equation\textquotedbl{}). Clarkson,
Clifton and February discussed the perturbations in the  
context of dust dominated Lemaitre-Tolman cosmologies  \cite{LTB}; their equations,
restricted to the FLRW metrics deformed by axial modes, coincide with
those of \cite{MW}. The new element of the present analysis, absent in \cite{MW} and \cite{LTB},
 is that  axial gravitational waves can interact with matter  and change its  azimuthal velocity ---
azimuthal velocity perturbations can be non-zero. 

The case with polar modes is discussed in Section 4. The calculation
was done from scratch, with the use of Mathematica.The full evolution
system consists of six equations; one can extract a part of the metric
that is described by the master equation, that turned out to have
exactly the same form as for the axial model and that agrees with
the corresponding result of \cite{LTB}. There was only one minor
difference in one of the remaining equations, comparing to \cite{LTB}; see a comment later on.

We should mention here an earlier analysis of Gundlach and Martin-Garcia \cite{GMG}
that uses approach  of Gerlach and Sengupta \cite{GS}, and that describes propagation of 
gravitational modes within time-dependent stellar enviroments. We understand that one  can get from these
the  cosmological (FLRW)  propagation equations --- see a discussion in \cite{LTB} --- 
but our analysis is independent.

Section 5 deals with the universe filled by perturbed perfect fluid.
We demonstrate that equations are consistent and we prove that   mass density 
and velocity (radial and poloidal components) of  the
material (perfect fluid) must be perturbed by a travelling polar gravitational wave
pulse.

The vacuum case, de Sitter spacetime, is considered in Section 6.
The whole system of 6 equations is consistent, reduces to 3 independent
equations, the master equation and two other conditions that completely
determine the linearized metric.

The next section shows that the Huygens principle is strictly valid
only in radiation spacetimes (with $\Lambda=0$), both for axial and
polar modes.

Final section gives a concise description of main results and comments
on their significance.

We assume standard gravitational units $c=8\pi G=1$.

\section{FLRW geometry and its perturbations }

We shall use standard coordinates, but with the conformal time coordinate;
the FLRW metric tensor reads

\begin{equation}
(g_{\mu\nu}^{(0)})=\left(\begin{array}{ccccc}
-a^{2}(\eta) & 0 & 0 & 0\\
0 & a^{2}(\eta) & 0 & 0\\
0 & 0 & a^{2}(\eta)f^{2} & 0\\
0 & 0 & 0 & a^{2}(\eta)f^{2}\sin^{2}\theta
\end{array}\right),
\end{equation}

\noindent where $f=f(r)=r,\sin r,\sinh r$ for the flat, closed, and
open universes, respectively. $d\Omega^{2}=d\theta^{2}+\sin^{2}\theta~d\phi^{2}$
is the line element on a unit sphere.

We shall use below the \textquotedbl{}conformal\textquotedbl{} Hubble
constant,

{ 
\begin{equation}
H(\eta):=\frac{\partial_{\eta}a}{a};
\end{equation}
}

it is related to the \textquotedbl{}ordinary\textquotedbl{} Hubble
constant ${\underline{H}}$ by ${\underline{H}}=\frac{H}{a}$.

\subsection{Metric Regge-Wheeler perturbations}

We shall employ results of Regge and Wheeler \cite{R-W}. Below we
put for notational simplicity $Y=Y(\theta):=Y_{l0}(\theta)$; thus

\begin{equation}
Y''=-l(l+1)Y-\cot\theta~Y',
\end{equation}

\noindent where $Y'=\partial_{\theta}Y$.

For completeness we shall consider both polar and axial modes; in
the latter case we again arrive at conclusions already known from
\cite{MW}.

\subsubsection{Axial modes}

The axially perturbed components of the metric read

\begin{equation}
g_{\mu\nu}=g_{\mu\nu}^{(0)}+e\cdot h_{\mu\nu}^{(o)}+o(e^{2}),
\end{equation}

\noindent where

\begin{equation}
(h_{\mu\nu}^{(o)})=\left(\begin{array}{ccccc}
0 & 0 & 0 & h_{0}\sin\theta~Y'\\
0 & 0 & 0 & h_{1}\sin\theta~Y'\\
0 & 0 & 0 & 0\\
 h_{0}\sin\theta~Y' & h_{1}\sin\theta~Y  & 0 & 0
\end{array}\right).\label{axialm}
\end{equation}

\noindent Here $h_{0}=h_{0}(\eta,r)$, $h_{1}=h_{1}(\eta,r)$. The
small parameter $e$ is introduced for convenience; it measures the
strength of perturbations.

\subsubsection{Polar modes }

The perturbed components of the metric corresponding to polar modes
are

\begin{equation}
g_{\mu\nu}=g_{\mu\nu}^{(0)}+e\cdot h_{\mu\nu}^{(p)}+o(e^{2}),
\end{equation}

\noindent where we used the convention of Clarkson, Clifton, and February
\cite{LTB}:

\begin{equation}
(h_{\mu\nu}^{(p)})=\left(\begin{array}{ccccc}
\left(\chi+\varphi\right)Y & \sigma Y & 0 & 0\\
 \sigma Y & \left(\chi+\varphi\right)Y & 0 & 0\\
0 & 0 & f^{2}\varphi Y & 0\\
0 & 0 & 0 & f^{2}\sin^{2}\theta~\varphi Y
\end{array}\right);
\end{equation}

\noindent as before, we do not write arguments of perturbing functions
explicitly. Thus $\chi=\chi(\eta,r)$, $\varphi=\varphi(\eta,r)$,
  $\sigma=\sigma(\eta,r)$.
  
 We need to say in this place, that the relevant perturbation matrix of \cite{LTB} contains an additional function $\psi $ in the metric 
 element $h_{00}$, but we checked that it must vanish for the FLRW background.

\subsection{The energy-momentum tensor}

In the former calculation concerning axial gravitational waves \cite{MW}
there was no need to perturb the cosmological background. This is
in contrast, as we shall see, with the polar modes. Below we shall
perturb the background material quantities in both cases, but finally
it appears that this may be avoided in the situation discussed in \cite{MW}.

The tensor $T_{\mu\nu}$ has the form:

\begin{equation}
T_{\mu\nu}=(\rho+p)u_{\mu}u_{\nu}+pg_{\mu\nu}-\Lambda g_{\mu\nu}.
\end{equation}

\noindent The mass density is given by

\begin{equation}
\rho=\rho_{0}\left(1+e\cdot\Delta(\eta,r)Y\right)+o(e^{2}),
\end{equation}

where $\rho_{0}$ is the background mass density and $e\cdot\Delta(\eta,r)Y$
is the mass density contrast.

The pressure splits onto $p_{0}$, the background pressure, and $e\cdot\Pi(\eta,r)Y$
--- the pressure contrast: 
\begin{equation}
p=p_{0}\left(1+e\cdot\Pi(\eta,r)Y\right)+o(e^{2}),
\end{equation}

The pressure and mass density perturbations $\Pi$ and $\Delta$ are
related, since $\rho$ and $p$ are related by an equation of state.

Finally, one should allow the possibility that matter is not necessarily
comoving with the unperturbed cosmological expansion. Thus the 4-velocity
of matter reads

\begin{eqnarray}
u_{0}&=&  \frac{2g_{00}^{(0)}+e\cdot h_{00}}{2a(\eta)}+o(e^{2}), \nonumber \\
u_1&=& e\cdot a(\eta)w(\eta,r)Y +o(e^{2}), \nonumber \\
u_2&= & e\cdot v(\eta,r)Y'+o(e^{2}), \nonumber \\
u_3&=&  e\cdot  \sin \theta \cdot u(\eta,r)Y'+o(e^{2})
\end{eqnarray}

This ensures that $u_{\mu}u^{\mu}=-1+o(e^{2})$.

\subsection{Background Friedman-Lemaitre solution}

The (background) isotropic and homogeneous solution of Friedman equations
satisfies following relations

\begin{equation}
\rho_{0}=\frac{3}{a^{2}}H^{2}+\frac{3}{a^{2}}\frac{1-f'^{2}}{f^{2}}-\Lambda,
\end{equation}
and

\begin{equation}
p_{0}=\Lambda-\frac{1}{a^{2}}\frac{1-f'^{2}}{f^{2}}-\frac{1}{a^{2}}H^{2}-\frac{2}{a^{2}}\dot{H}.
\end{equation}

From these two equations one arrives at

\begin{equation}
\frac{a^{2}}{2}\left(\frac{1}{3}\rho_{0}-p_{0}+\frac{4}{3}\Lambda\right)=H^{2}+\dot{H}+\frac{1-f'^{2}}{f^{2}};\label{wyraz_przez_rho_p_i_Lambda}
\end{equation}

this relation is used later.

\section{Main calculations: axial modes}

Linearized Einstein equations corresponding to the metric (\ref{axialm}),
read

\begin{equation}
\Delta\cdot\rho_{0}=0\label{mody_osiowe_Delta}
\end{equation}

\begin{equation}
\Pi\cdot p_{0}=0\label{mody_osiowe_Pi}
\end{equation}

\begin{equation}
w\cdot\left(\rho_{0}+p_{0}\right)=0\label{mody_osiowe_w}
\end{equation}

\begin{equation}
v\cdot\left(\rho_{0}+p_{0}\right)=0\label{mody_osiowe_v}
\end{equation}

\begin{equation}
h'_{1}=\dot{h}_{0}\label{mody_osiowe_1}
\end{equation}

\begin{eqnarray}
 && \dot{h}'_{1}-h''_{0}-2Hh'_{1}+2\frac{f'}{f}\dot{h}_{1}-4H\frac{f'}{f}h_{1}+ \nonumber \\
 && +\frac{l(l+1)+4f'^{2}-4}{f^{2}}h_{0}=-2a^3 \left( \rho_0 +p_0\right) u \label{mody_osiowe_2}
 \end{eqnarray}

\begin{equation}
\ddot{h}_{1}-\dot{h}'_{0}-2H\dot{h}_{1}+2\frac{f'}{f}\dot{h}_{0}-2\dot{H}h_{1}+\frac{l(l+1)-2}{f^{2}}h_{1}=0\label{mody_osiowe_3}
\end{equation}

The first four equations immediately imply $\Delta=\Pi=w=v=0$.    In what follows we form the 
following linear combination of Eqs. (\ref{mody_osiowe_1})) ---- (\ref{mody_osiowe_3}))

\begin{eqnarray}
&&\partial_{r}(f^{2}\cdot(\ref{mody_osiowe_3}))-\partial_{\eta}(f^{2}\cdot(\ref{mody_osiowe_2}))-(l(l+1)-2)\cdot(\ref{mody_osiowe_1})=\nonumber\\
&&f^2\partial_\eta \bigl( 2a^3 \left( \rho_0 +p_0\right) u \bigr) .
\label{u}
\end{eqnarray}

The expression $\partial_{r}(f^{2}\cdot(\ref{mody_osiowe_3}))$ means
that the left hand side of equation no (\ref{mody_osiowe_3}) is multiplied
by $f^{2}$ and then differentiated with respect $r$. The same convention
is applied in the remaining part of this paper. 

It appears that the left hand side  of (\ref{u}) vanishes  identially, which in turn implies that the angular perturbing term $u$ acquires following form:

\begin{equation}
u= \frac{C(r)}{a^3 \left( \rho_0 +p_0\right)}.
\label{angular}
\end{equation}

One may choose  $C(r)=0$ --- that was implicitly assumed  in  \cite{MW}) and \cite{LTB}.
Then the material content of the background cosmological  spacetime
does not feel the propagation of axial modes. In the generic case initial data for axial modes yield 
nonzero $C(r)$ --- see Eq. (\ref{mody_osiowe_2}). Therefore a pulse of an axial radiation enforces infinitesimal rotation of a cosmological fluid around $z-$axis.  One may expect, judging from the form of the right hand side of (\ref{angular}),
  that this perturbation increases, $u\propto a$, during radiation era and remains constant during dust-dominated era.
 This issue requires a more careful analysis that will be done elsewhere \cite{WK_EM}.

Furthermore, inserting $(\ref{mody_osiowe_1})$ into $(\ref{mody_osiowe_3})$,
one arrives at

\begin{equation}
\ddot{h}_{1}-h''_{1}-2H\dot{h}_{1}+2\frac{f'}{f}h'_{1}-2\dot{H}h_{1}+\frac{l(l+1)-2}{f^{2}}h_{1}=0.\label{mody_osiowe_falowe}
\end{equation}

Defining now a new quantity $Q(\eta,r)$ by

\begin{equation}
h_{1}(\eta,r)=f(r)a(\eta)Q(\eta,r),
\end{equation}

\noindent and using (\ref{wyraz_przez_rho_p_i_Lambda}), one arrives
at a single wave equation

\begin{equation}
\ddot{Q}-Q''+\frac{l(l+1)}{f^{2}}Q-\frac{1}{2}a^{2}\left(\frac{1}{3}\rho_{0}-p_{0}+\frac{4}{3}\Lambda\right)Q=0.\label{mody_osiowe_falowe_na_Q}
\end{equation}

This agrees with the corresponding result of \cite{MW} and \cite{LTB}.

The master equation (\ref{mody_osiowe_falowe_na_Q}) can be solved
independently of the remaining equations --- it constitutes one of
the two independent gravitational modes.

From Eq. (\ref{mody_osiowe_1}) one gets $h_{0}$:

\begin{equation}
h_{0}(\eta,r)=A(r)+\int\limits _{\eta_0}^{\eta}h'_{1}(\tau,r)d\tau.\label{mody_osiowe_h0}
\end{equation}

The function $A(r)$ is arbitrary, but if $h_0(\eta,r)$ vanishes
at the initial hypersurface, then $A(r)=0$.

\section{Polar modes: equations}

The linearization of Einstein equation yields, after straightforward
calculations, three groups of equations.

Two equations describe the evolution of the mass density contrast

\[
\Delta(\eta,r)=\frac{1}{2a^{4}\rho_{0}}\left[\left(\frac{l(l+1)+6f'^{2}-4}{f^{2}}+2H^{2}\right)\chi+\right.
\]
\begin{equation}
+2\left(\frac{l(l+1)+3f'^{2}-3}{f^{2}}-3H^{2}\right)\varphi-\label{mody_polarne_Delta}
\end{equation}
\[
\left.-8H\frac{f'}{f}\sigma+2\frac{f'}{f}(\chi'-2\varphi')+2H(\dot{\chi}+3\dot{\varphi}-2\sigma')-2\varphi''\right],
\]

and the pressure contrast,

\[
\Pi(\eta,r)=\frac{1}{2a^{4}p_{0}}\left[\left(\frac{l(l+1)-2f'^{2}}{f^{2}}+2H^{2}-4\dot{H}\right)\chi+\right.
\]
\begin{equation}
\left.2\left(\frac{1-f'^{2}}{f^{2}}+H^{2}\right)\varphi+2\frac{f'}{f}\chi'+2H(\dot{\varphi}-\dot{\chi})-2\ddot{\varphi}\right].\label{mody_polarne_Pi}
\end{equation}

The second group describes deformation of the 2 components of the
4-velocity:

\[
w(\eta,r)=\frac{1}{2a^{4}(\rho_{0}+p_{0})}\left[4H\frac{f'}{f}\chi-\frac{l(l+1)+4f'^{2}-4}{f^{2}}\sigma-\right.
\]

\begin{equation}
\left.-2\frac{f'}{f}\dot{\chi}+2H(\chi'-\varphi')+2\dot{\varphi}'\right];\label{mody_polarne_w}
\end{equation}

\begin{equation}
v(\eta,r)=\frac{-2H\varphi-\sigma'+2\dot{\varphi}+\dot{\chi}}{2a^{3}(\rho_{0}+p_{0})}\label{mody_polarne_v}
\end{equation}

The last pair of equations does not depend explicitly on matter:

\begin{equation}
\chi'=\dot{\sigma}\label{mody_polarne_1}
\end{equation}

\begin{equation}
\ddot{\chi}-\chi''-2H\dot{\chi}+2\frac{f'}{f}\chi'-2\dot{H}\chi+\frac{l(l+1)-2}{f^{2}}\chi=0\label{mody_polarne_2}
\end{equation}

Notice that the equation (\ref{mody_polarne_2}) onto the function
$\chi$ separates from the rest. With the substitution

\begin{equation}
\chi(\eta,r)=f(r)a(\eta)Q(\eta,r),
\end{equation}

\noindent one arrives at a result that coincides with the equation
(\ref{mody_osiowe_falowe_na_Q}).

Again it appears that the master equation can be solved independently
of the remaining equations. This implies that the metric functions
$\chi$ constitute one gravitational degree of freedom. Moreover,
the evolution equations of axial and polar modes have the same form.
The remaining metric functions, $\sigma$ and $\varphi$ are not arbitrary.
For instance one has

\begin{equation}
\sigma(\eta,r)=B(r)+\int\limits _{\eta_{0}}^{\eta}\chi'(\tau,r)d\tau;\label{mody_osiowe_sigma}
\end{equation}

the function $B(r)$ that appears here is arbitrary, but if one assumes
that initially $\sigma(\eta,r)$ vanishes, then $B(r)=0$.

  One of   our equations that describes  behaviour of the radial velocity perturbation,  Eq. no. (\ref{mody_polarne_w}), disagrees with the Friedmannian limit 
 of Eq. (44)  in \cite{LTB}; in order to achieve agreement, there should be $8\pi \rho $  instead of $4\pi \rho $ in Eq. (44). Below we show the (wrong) equation implied by Eq. (44): 
\[
w(\eta,r)=\frac{1}{2a^{4}(\rho_{0}+p_{0})}\left[4H\frac{f'}{f}\chi-\frac{l(l+1)+4f'^2-4 }{f^{2}}\sigma-\right.
\]

\begin{equation}
\left. +3 \sigma (\frac{1-f'^2}{f^2}+H^2) -2\frac{f'}{f}\dot{\chi}+2H(\chi'-\varphi')+2\dot{\varphi}'\right];\label{mody_polarne_wF}
\end{equation}
   
 the spurious term  is $ 3 \sigma (\frac{1-f'^2}{f^2}+H^2)$.
   
We checked that Eqs (\ref{mody_polarne_1}), (\ref{mody_polarne_2})  and (\ref{mody_polarne_phi}) (see the next section) agree with 
corresponding equations of \cite{GMG}.

\section{Polar perturbations and material inhomogeneities}

\subsection{Metric and material perturbations}

The part of the perturbing metric that contains the function $\varphi$
represents a conformally flat perturbation. The evolution of $\varphi$
is influenced by other metric polar perturbation functions - $\sigma$
and $\chi$. We show that the full set of equations is consistent
and fully describes the propagation of polar modes and the creation
of inhomogeneities in the universe filled by perfect fluid.

Assume $p=K\rho^{\Gamma}$ as the equation of state. Then the speed
of sound $c_{s}$ is given by $c_{s}^{2}=\partial_{\rho}p$ and $\Delta\rho_{0}=\Pi p_{0}/c_{s}^{2}$.
Multiply both sides of Eq. (\ref{mody_polarne_Pi}) by $2a^{4}p_{0}/c_{s}^{2}$,
and Eq. (\ref{mody_polarne_Delta}) by $2a^{4}\rho_{0}$ and subtract
the obtained equations. After an easy calculation one arrives at

\begin{eqnarray}
 &  & \frac{2}{c_{s}^{2}}\ddot{\varphi}-2\varphi''-4\varphi'\frac{f'}{f}+\nonumber \\
 &  & +\left(\frac{l\left(l+1\right)\frac{c_{s}^{2}-1}{c_{s}^{2}}+2f'^{2}\frac{3c_{s}^{2}+1}{c_{s}^{2}}-4}{f^{2}}+2\frac{c_{s}^{2}-1}{c_{s}^{2}}H^{2}+\frac{4}{c_{s}^{2}}\dot{H}\right)\chi+\nonumber \\
 &  & +2\left(\frac{l\left(l+1\right)+\left(f'^{2}-1\right)\frac{3c_{s}^{2}+1}{c_{s}^{2}}}{f^{2}}-\frac{3c_{s}^{2}+1}{c_{s}^{2}}H^{2}\right)\varphi-\nonumber \\
 &  & -4H\left(\sigma'+2\frac{f'}{f}\sigma\right)+2\frac{f'}{f}\frac{c_{s}^{2}-1}{c_{s}^{2}}\chi'+\nonumber \\
 &  & +2H\left(\dot{\chi}\frac{1+c_{s}^{2}}{c_{s}^{2}}+\dot{\varphi}\frac{3c_{s}^{2}-1}{c_{s}^{2}}\right)=0\label{mody_polarne_phi}
\end{eqnarray}

This is a hyperbolic differential equation, whose initial data ---
$\varphi$ and $\dot{\varphi}$ --- are dictated by initial values
of the mass density contrast $\Delta$ and the velocity ($v$) perturbation.
Indeed, one finds from Eq. (\ref{mody_polarne_v})

\begin{equation}
2\dot{\varphi}=2v(\eta,r)a^{3}(\rho_{0}+p_{0})+2H\varphi+\sigma'-\dot{\chi};\label{mody_polarne_vv}
\end{equation}

inserting that into Eq. (\ref{mody_polarne_Delta}) yields an ordinary
differential equation of second order onto the unknown $\varphi$.
All other data are known from the background geometry ($H$ and $a$)
or from solving of the master equation ($\chi,\dot{\chi},\chi'$ and
$\sigma,\sigma'$). The appropriate boundary conditions are $\varphi'(r=0)=0$
and $\varphi(r=\infty)=0$. If in addition the initial mass density
contrast $\Delta$ is prescribed, then the initial values of $\varphi$
can be determined uniquely. The initial \textquotedbl{}velocity\textquotedbl{}
$\dot{\varphi}$ is then obtained from the initial velocity perturbation
dictated by (\ref{mody_polarne_vv}). These initial data for the equation
(\ref{mody_polarne_phi}) give rise to an evolving solution $\varphi$;
 this in turn, determines in later times the mass density contrast
$\Delta$ and the pressure contrast $\Pi$.

Deformations of the velocity components $w(\eta,r)$ and $v(\eta,r)$
are dictated by equations (\ref{mody_polarne_w}) and (\ref{mody_polarne_v}).

We would like to point out that the above reasoning proves consistency
of equations. It suggests also that even if initially density and
velocity perturbations do vanish, then they are expected to appear
in later times. Next subsection proves a stronger statement.

\subsection{Polar waves must influence cosmological matter}

In the remainder of this Section we show, \textbf{that if a cosmological
spacetime is nonvacuous, $p_{0}\ge0$ and $\varrho_{0}>0$, then propagating
polar waves must disturb the material medium}. We shall use the method
of contradiction --- assume that matter perturbations are absent, i.e.  $\Pi=\Delta=w=v\equiv0$ and derive from that $\sigma=\chi=\varphi\equiv0$ --- i. e., that polar modes
are absent. {\bf This means that nonzero  polar modes induce matter disturbances.}

The propagation equations $(\ref{mody_polarne_Delta})-(\ref{mody_polarne_2})$
read, assuming $\Pi=\Delta=w=v\equiv0$:

\[
\left(\frac{l(l+1)+6f'^{2}-4}{f^{2}}+2H^{2} \right)\chi+
\]
\[
+2\left(\frac{l(l+1)+3f'^{2}-3}{f^{2}}-3H^{2} \right)\varphi-
\]
\begin{equation}
-8H\frac{f'}{f}\sigma+2\frac{f'}{f}(\chi'-2\varphi')+2H(\dot{\chi}+3\dot{\varphi}-2\sigma')-2\varphi''=0\label{sprzecznosc_1}
\end{equation}

\[
\left(\frac{l(l+1)-2f'^{2}}{f^{2}}+2H^{2} -4\dot H\right)\chi+
\]
\begin{equation}
+2\left(\frac{1-f'^{2}}{f^{2}}+H^{2}\right)\varphi+2\frac{f'}{f}\chi'+2H(\dot{\varphi}-\dot{\chi})-2\ddot{\varphi}=0\label{sprzecznosc_2}
\end{equation}

\begin{equation}
\frac{l(l+1)+4f'^{2}-4}{f^{2}}\sigma-4H\frac{f'}{f}\chi+2\frac{f'}{f}\dot{\chi}-2H(\chi'-\varphi')-2\dot{\varphi}'=0\label{sprzecznosc_3}
\end{equation}

\begin{equation}
2H\varphi+\sigma'-2\dot{\varphi}-\dot{\chi}=0\label{sprzecznosc_4}
\end{equation}

\begin{equation}
\chi'-\dot{\sigma}=0\label{sprzecznosc_5}
\end{equation}

\begin{equation}
\ddot{\chi}-\chi''-2H\dot{\chi}+2\frac{f'}{f}\chi'-2H'\chi+\frac{l(l+1)-2}{f^{2}}\chi=0\label{sprzecznosc_6}
\end{equation}

It is easy to show

\textbf{Lemma 1.} If $\rho_{0}+p_{0}>0$ then $\varphi(\eta,r)=-\chi(\eta,r)$.

Indeed, performing on Eqs. (\ref{sprzecznosc_2}), (\ref{sprzecznosc_4})
--- (\ref{sprzecznosc_6}) the following operations $(\ref{sprzecznosc_2})-(\ref{sprzecznosc_6})-\partial_{r}(\ref{sprzecznosc_5})-\partial_{\eta}(\ref{sprzecznosc_4})$,
one arrives at

\begin{equation}
a^{2}(\rho_{0}+p_{0})(\varphi+\chi)=0.
\end{equation}

The set of the foregoing equations reduces now to the following system

\begin{equation}
\left(\frac{2-l(l+1)}{f^{2}}+8H^{2}\right)\chi-8H\frac{f'}{f}\sigma+6\frac{f'}{f}\chi'-4H(\dot{\chi}+\sigma')+2\chi''=0\label{Asprzecznosc_1}
\end{equation}

\begin{equation}
\left(\frac{l(l+1)-2}{f^{2}}-4\dot{H}\right)\chi+2\frac{f'}{f}\chi'-4H\dot{\chi}+2\ddot{\chi}=0\label{Asprzecznosc_2}
\end{equation}

\begin{equation}
\frac{l(l+1)+4f'^{2}-4}{f^{2}}\sigma-4H\frac{f'}{f}\chi+2\frac{f'}{f}\dot{\chi}-4H\chi'+2\dot{\chi}'=0\label{Asprzecznosc_3}
\end{equation}

\begin{equation}
\sigma'+\dot{\chi}-2H\chi=0\label{Asprzecznosc_4}
\end{equation}

\begin{equation}
\chi'-\dot{\sigma}=0\label{Asprzecznosc_5}
\end{equation}

The combination of operations $(\ref{Asprzecznosc_2})-2(\partial_{\eta}(\ref{Asprzecznosc_4})+\partial_{r}(\ref{Asprzecznosc_5}))$
yields

\begin{equation}
0=\frac{l(l+1)-2}{f^{2}}\chi+2\frac{f'}{f}\chi'-2\chi''.\label{AaAsprzecznosc_2}
\end{equation}

This equation can be written as

\begin{equation}
0=\frac{l(l+1)-2}{2f^{3}}\chi-\frac{d}{dr}\frac{\chi'}{f}.\label{AaAsprzecznosc_3}
\end{equation}

We show that

\textbf{Lemma 2}. Eq. (\ref{AaAsprzecznosc_3}) does not possess nonzero
solutions of compact support.

Proof.

Indeed, let there exists two closest points $r_{1}$ and $r_{2}$
($r_{1}<r_{2}$) such that $\chi(r_{1})=\chi(r_{2})=0$, and integrate
(\ref{AaAsprzecznosc_3}) over the interval $(r_{1},r_{2})$. We arrive
at 
\begin{equation}
0=\int_{r_{1}}^{r_{2}}dr\frac{l(l+1)-2}{2f^{3}}\chi-\frac{\chi'(r_{2})}{f(r_{2})}+\frac{\chi'(r_{1})}{f(r_{1})}.\label{AaAsprzecznosc_4}
\end{equation}
Let $\chi$ be strictly positive in the open interval $(r_{1},r_{2})$.
Then necessarily $\frac{\chi'(r_{2})}{f(r_{2})}>>\frac{\chi'(r_{1})}{f(r_{1})}$;
but that means that $\chi$ is monotonically increasing, which contradicts
the assumption that $\chi(r_{2})$ vanishes. In the same manner one
considers the case with $\chi$ being strictly negative in the open
interval $(r_{1},r_{2})$. That ends the proof of the Lemma 2.

This allows us to conclude, that the vanishing of material cosmological
perturbations $\Pi=\Delta=w=v\equiv0$ enforces the absence of polar
gravitational waves, $\sigma=\chi=\varphi\equiv0$.

This in turn means that travelling gravitational polar pulses must
leave inhomogeneous and nonisotropic tracks in the cosmological background.

\section{ Polar modes in the de Sitter universe}

In the de Sitter universe matter is absent, $\rho_{0}=p_{0}=0$.  S. Viaggiu has shown, using Laplace transforms, 
that polar perturbations can also be expressed in terms of four independent 
integrable differential equations \cite{V}. We
have  six equations describing the propagation of gravitational
waves, while there are only two free initial data, corresponding to
the one degree of freedom carried by polar modes. We show in this
section that these equations are consistent and that the whole dynamics
is described by the master equation.

The evolution equation (\ref{wyraz_przez_rho_p_i_Lambda}) describing
the background geometry reads now 
\begin{equation}
\frac{1}{3}a^{2}\Lambda=H^{2}+\frac{1-f'^{2}}{f^{2}}=\dot{H}.\label{brak_materii_warunek}
\end{equation}

It is useful to notice that $\frac{d}{dr}\frac{1-f'^{2}}{f^{2}}=0$,
that is 
\begin{equation}
f''+\frac{1-(f')^{2}}{f}=0.\label{f}
\end{equation}
This identity can be derived from (\ref{brak_materii_warunek}) ---
notice that $H$ depends only on $\eta$ --- or directly from the
definition of $f$.

Equations $(\ref{mody_polarne_Delta})-(\ref{mody_polarne_2})$ take
following form:

\[
\left(\frac{l(l+1)+6f'^{2}-4}{f^{2}}+2H^{2}\right)\chi+
\]
\[
2\left(\frac{l(l+1)+3f'^{2}-3}{f^{2}}-3H^{2}\right)\varphi-8H\frac{f'}{f}\sigma+
\]
\begin{equation}
+2\frac{f'}{f}(\chi'-2\varphi')+2H(\dot{\chi}+3\dot{\varphi}-2\sigma')-2\varphi''=0\label{brak_materii_1}
\end{equation}

\[
\left(\frac{l(l+1)+2f'^{2}-4}{f^{2}}-2H^{2}\right)\chi+2\left(\frac{1-f'^{2}}{f^{2}}+H^{2}\right)\varphi
\]
\begin{equation}
+2\frac{f'}{f}\chi'+2H(\dot{\varphi}-\dot{\chi})-2\ddot{\varphi}=0\label{brak_materii_2}
\end{equation}

\begin{equation}
4H\frac{f'}{f}\chi-\frac{l(l+1)+4f'^{2}-4}{f^{2}}\sigma-2\frac{f'}{f}\dot{\chi}+2H(\chi'-\varphi')+2\dot{\varphi}'=0\label{brak_materii_3}
\end{equation}

\begin{equation}
2(\dot{\varphi}-H\varphi)-\sigma'+\dot{\chi}=0\label{brak_materii_4}
\end{equation}

\begin{equation}
\chi'-\dot{\sigma}=0\label{brak_materii_5}
\end{equation}

\begin{equation}
\ddot{\chi}-\chi''-2H\dot{\chi}+2\frac{f'}{f}\chi'+\left(\frac{l(l+1)+2f'^{2}-4}{f^{2}}-2H^{2}\right)\chi=0\label{brak_materii_6}
\end{equation}

In the first step we shall express equations (\ref{brak_materii_1}),
(\ref{brak_materii_2}) and (\ref{brak_materii_3}) as linear combinations
of the remaining three equations (\ref{brak_materii_4}) --- (\ref{brak_materii_6}).
We assume that at some initial time $\eta_{0}$ the left hand side
of Eqs. (\ref{brak_materii_3}) does vanish.

One can check by straightforward calculation that

\begin{equation}
(\ref{brak_materii_2})=(\ref{brak_materii_6})+\partial_{r}(\ref{brak_materii_5})-\partial_{\eta}(\ref{brak_materii_4});
\end{equation}
in the course of calculations we replaced $\dot{H}$, using Eq. (\ref{brak_materii_warunek}).

Direct calculation shows that

\begin{equation}
\partial_{\eta}(\ref{brak_materii_3})=-\frac{{l(l+1)+4f'^{2}-4}}{f^{2}}(\ref{brak_materii_5})+\partial_{r}(\ref{brak_materii_2})+\frac{2f'}{f}(\ref{brak_materii_6})
\end{equation}
Here we have to use the identity (\ref{f}).

It is clear that if (\ref{brak_materii_4}), (\ref{brak_materii_5})
and (\ref{brak_materii_6}) are satisfied, then equation (\ref{brak_materii_2})
also holds. Eq. (\ref{brak_materii_3}) is also satisfied modulo a
time-independent function, say $f_{3}(r)$; but now we use the assumption
that initially the velocity perturbation vanishes; thus $f_{3}(r)=0$
and (\ref{brak_materii_3}) is also valid.

The last equation, (\ref{brak_materii_1}), satisfies following relation:

\[
-a\partial_{\eta}\left((\ref{brak_materii_1})\frac{1}{a}\right)=H(\ref{brak_materii_6})-H\partial_{r}(\ref{brak_materii_5})-3H\partial_{\eta}(\ref{brak_materii_4})-~~~~~~~~~~~~~~~~~~
\]
\begin{equation}
~~~~~~~~~~-8H\frac{f'}{f}(\ref{brak_materii_5})+2\frac{f'}{f}(\ref{brak_materii_3})-\frac{l(l+1)}{f^{2}}(\ref{brak_materii_4})+\partial_{r}(\ref{brak_materii_3})\label{eq:brak_materii_r2}
\end{equation}

We know, from the preceding analysis, that the right hand side of
(\ref{eq:brak_materii_r2}) vanishes. Therefore (\ref{brak_materii_1})
holds true modulo a time-independent function $f_{1}(r)$; but $f_{1}(r)=0$,
since we assumed that the mass density and its perturbation are absent.
That means that (\ref{brak_materii_1}) is valid. In summary, this
reasoning shows that the only relevant equations are (\ref{brak_materii_4})
--- (\ref{brak_materii_6}).

In the second step one shows that metric functions $\varphi$ and
$\sigma$ can be derived from Eqs. (\ref{brak_materii_4}) and (\ref{brak_materii_5}).
They become functionally dependent on solutions $\chi$ of the master
equation --- see explicit formulae in preceding sections --- modulo
a time-independent function; but the latter can be set to zero, with
appropriate initial conditions. Thus the whole dynamics reduces to
the single master equation, as in the analysis of Zerilli in the Schwarzschild
spacetime \cite{Zerilli}.

\section{The Huygens principle}

We adopt the following version of the Huygens principle. Purely outgoing
compact pulses of radiation, that at $\eta_0$ occupy a shell $r_{in}\le r\le r_{out}$, have to move within space-time regions bounded by a pair of
outgoing null cones $\eta=r+const$. They should not leave any tails
trailing behind the main pulse. Analogous property is obeyed by purely
ingoing waves. It is well known that solutions of the massless scalar field equation  
$\box \Phi =0$ satisfy the above Huygens principle in $1+3$ dimensional Minkowski spacetime, 
while massive scalar fields do not obey it. Radiation shocks --- with initial data located strictly
on a sphere --- should move strictly along null cones. Various versions
are discussed in the monograph of Hadamard \cite{Hadamard}.

During the radiation epoch and assuming that $\Lambda=0$, the Huygens
principle holds. Indeed, the form of the master equation describing
the propagation of axial and polar waves 
\begin{equation}
\ddot{Q}-Q''+\frac{l(l+1)}{f^{2}}Q=0,\label{fala_spelniajaca_zasade_Huygensa}
\end{equation}

\noindent where $Q=Q(\eta,r)$ is exactly the same as that describing
the propagation of electromagnetic fields in FLRW spacetimes. Therefore
its general solution has the form (see, for instance, \cite{MWK})
\begin{equation}
\phi_{l}(r,\eta)=f^{l}\,\underbrace{{\partial_{r}}\,{\frac{1}{f}}\,{\partial_{r}}\,{\frac{1}{f}}\cdots{\partial_{r}}\,\bigl(\frac{g+h}{f}}_{l}\bigr)\label{2.2}
\end{equation}
where functions $h$ and $g$ depend on the combinations $r-\eta$ or
$r+\eta$, respectively. The multipole index $l$ exceeds 1 for the
wave-like solution. This solution obviously satisfies the Huygens
principle. Another argument for the validity of the Huygens principle
employs the conformal flatness of the FLRW models and conformal invariance
of the Maxwell equations in vacuum.

We should point that there exist other forms of explicit solutions
of Eq. (\ref{fala_spelniajaca_zasade_Huygensa}) \cite{Friedlander}.

In all remaining cases --- when the dark energy ($\Lambda\ne0$),
and/or nonradiational matter are present --- the Huygens principle
is broken. That means that tail terms --- delayed signals tracking
the main wave pulse --- should be present. This effect depends, however,
on the ratio of the length-wave to the Hubble radius and it is weak
when this ratio is small (\cite{MW} and \cite{MWK}).  Therefore it should not be observed on
subhorizon scales. This in turn
implies that it cannot be observed in present or presently planned
gravitational waves detectors.

\section{Summary}

It was not our intention to focus on the derivation of equations describing
propagation of gravitational waves in cosmological spacetimes. It
appeared, however, that there is no full consensus as to the form of these
equations. We have got almost the same results as in \cite{LTB} (after
specialization to dust-like FLRW cosmologies) --- in particular "master equations"
are the same and there is only one minor difference in one of the
remaining equations that describes velocity perturbations. 
Take a note, however, that Clarkson et al. assume dust while
we deal with more general polytropic fluids.  Our derivation  is completely independent of
the analysis of Gundlach and Martin-Garcia \cite{GMG}, but our main equations agree with those 3 equations
describing polar modes (including the master equation) that they give explicitly.
 
One of the results of this paper is the proof that polar gravitational
modes lead inevitably --- even in the linearized approximation ---
to matter inhomogeneities and anisotropies. These latter must emerge
in later evolution, even if matter is distributed isotropically and
homogeneously at the initial hypersurface. Polar gravitational waves perturb both density and
non-azimuthal components  of the velocity.  We would like to point
out that this conclusion is consistent with earlier studies \cite{1,13,15}
and also with the investigation of C. ~Clarkson, T.~Clifton and S.~February
\cite{LTB}.  We show also that axial gravitational waves can perturb --- in the linear 
approximation --- only azimuthal  velocity of cosmological fluid, leaving intact its density.
 It is not clear to us whether these properties of axial and polar modes would 
 have observational consequences, but in principle they would, also in subhorizon scales.
 
 The Huygens principle is strictly obeyed by polar and axial modes
during the radiation epoch and broken in other cosmological epochs.
This breakdown effect is small when the ratio of the length-wave to
the Hubble radius is small (\cite{MW}, \cite{praca doktorska}, \cite{MWK}).   Therefore it is  of no
significance in local astronomical observations, but it may manifest on superhorizon
scales. Primordial gravitational waves can be produced \cite{NSK} during formation of 
putative primordial black holes \cite{ZN}; whether they   may
leave detectable imprints in the cosmic microwave background radiation  is an open question
\cite{K}.

\end{document}